\documentstyle[12pt,epsf]{article}

\begin{document}

{\hfill AS-ITP 95-7}

\begin{center}
{\large \bf $b \to s \gamma$ Decay and Right-handed Top-bottom Charged Current}

{\bf  Cai-Dian L\"{u}$^{a,b}$\footnote{ E-mail: lucd@itp.ac.cn.},
 Jing-Liang Hu$^c$ and Chongshou Gao$^{a,b,d}$ \\
a CCAST(World Laboratory), P.O.Box 8730, Beijing 100080, China\footnote{Not
mailing address.}\\
b Institute of Theoretical Physics, Academia Sinica, P.O.Box 2735,\\
Beijing 100080, China \\
c Institute of High Energy Physics, Academia Sinica, P.O.Box 918,\\
Beijing 100039, China \\
d Physics Department, Peking University, Beijing 100871, China}

{\today}
\end{center}

\begin{abstract}
We introduce an anomalous top quark coupling (right-handed current) into
Standard Model Lagrangian. Based on this, a more complete calculation
of $b \to s\gamma $ decay
including leading log QCD corrections from $m_{top}$
to $M_W$ in addition to corrections  from $M_{W}$ to $m_b$ is given.
The inclusive decay rate is found to be suppressed comparing with the
case without QCD running from $m_t$ to $M_W$ except at the time of
small values of $|f_R^{tb}|$. e.g. when $f_R^{tb}=-0.08$, it is only
$1/10$ of the value given before. As $|f_R^{tb}|$ goes smaller, this
contribution is an enhancement like standard model case.
 From the newly experiment of CLEO Collaboration, strict restrictions to
parameters of this top-bottom quark coupling are found.
\end{abstract}
\bigskip


\newpage

		\section{Introduction}

The standard model(SM)  has achieved great success recent years.
However, there is still a vast interest beyond standard model.
It is well known that the process $b \to s\gamma $ is extremely sensitive
to new physics beyond the Standard Model\cite{Hew1}.
It has been argued that this experiment provides more
information about restrictions on the Standard Model, 2-Higgs
doublet model, Supersymmetry, Technicolor and etc.
Since the top quark is much heavier than other fermions, and its
interactions may be quite sensitive to new physics, the interactions
of the top quark are of special importance. In ref.\cite{fuj}, a
right-handed coupling of the top-bottom charged current is added
to the standard model(SM).  Based on this,
the authors give out the constraints to right-handed top quark current
by $b\to s \gamma$ decay. In fact, the right handed current is also
a low energy phenomena of
$SU(2)_L \times SU(2)_R \times U(1)$ left-right symmetric
model\cite{lr}. In this model, the right handed charged current is
induced by the $W_L-W_R$ mixing in addition to the usual left-handed
current in the standard model. Interference of the right- and left-handed
currents in the penguin diagram enables a chirality-flip by the top
quark mass $m_t$ inside the loop, and therefore leads to an amplitude
proportional to $m_t$\cite{coc,cho1,bab}. (In standard model,
the amplitude is proportional to the mass of the bottom or strange quark,
because the pure (V-A) structure of the charged currents requires the
chirality-flip to
proceed only through the mass of the initial or the final state quark.)
Therefore, large contributions from top-bottom quark right-handed current
to $b \to s \gamma$ amplitude occur. To first order approximation,
this just like to add a right-handed current to top-bottom quark coupling
in SM.

Recently the CLEO Collaboration has measured the inclusive branching
ratio of $b \to s \gamma$ to be\cite{cleo2}
\begin{equation}
Br( b \to s \gamma )=(2.32\pm 0.51 \pm 0.29 \pm 0.32) \times 10 ^{-4}.
\end{equation}
Corresponding to 95\% confidence level, the range is
$1 \times 10^{-4} < Br(b \to s \gamma )<4\times 10 ^{-4}$. This is a more
 stringent constraints compared with previous CLEO experiments\cite{cleo1}
cited by the above papers\cite{fuj,cho1,bab}. Furthermore, with more precise
experiments, a more accurate
theoretical calculation of this decay rate is also needed.

The inclusive $b\to s\gamma$ decay rate is often assumed to be well described
by the spectator model, where the b quark goes a radiative decay. The QCD
corrections to this decay have been calculated to leading logarithmic
accuracy by many authors\cite{Grin,Grig,Cel,Mis,Yao,Ciu}, and are known
to enhance the decay rate within SM by a factor of 3-4. This enhancement,
however, is subject to large uncertainties due to many reasons\cite{Bur}.
One of these reasons is the QCD running from top quark scale to W boson
scale, which has been
discussed in SM by ref.\cite{Cho,lcd1}, in 2-Higgs doublet model by
ref.\cite{lcd2}, and in Supersymmetry case by ref.\cite{Anl}. This
contribution which is usually considered as a next-leading order effect
was found to give an additional enhancement up to 20\% in the case of a
 much heavy top quark mass.

In our present paper, by introducing a right-handed charged current
to top-bottom couplings, we recalculate the $b \to s \gamma$ decay
including QCD running from $m_{top}$ to $M_W$, in addition to corrections
from $M_W$ to $m_b$, in order to give a more complete leading log result
in this model. Since the branching ratio is proportional to $|f_R^{tb}
m_t/m_b|^2$, it is more complicated than that in SM or other models.
Finally more recent CLEO experiment is used to give
more precise restrictions to this anomalous top quark coupling.

In the next section, we first integrate out the top quark,
generating an effective five-quark theory. By using the
renormalization group equation, we run the effective field theory
down to the W-scale where the weak bosons are removed. Then
we continue running the effective field theory
down to b-quark scale to include QCD corrections from $M_W$ to $m_b$.
In section 3, the rate of radiative $b$ decay is obtained.
Section 4 is a short summary.

 \section{QCD Corrections to $b \to s \gamma$ decay }

In standard model and many other models, there is no flavor changing
neutral current at tree level. It is only occurred through electroweak
 loop. In standard model Lagrangian, the relevant charged current in
$R_{\xi}$ gauge reads
\begin{eqnarray}
 {\cal L}_{CC}&= &
  \frac{1}{\sqrt{2}} \mu^{\epsilon/2} g_2\left(\begin{array}{ccc}
    \overline{u} & \overline{c} & \overline{t} \end{array}\right)_L
\gamma_{\mu} V\left(\begin{array}{c} d \\ s \\ b \end{array}\right)_L
	  W_+^{\mu} \nonumber\\
        &  +&\frac{1}{\sqrt{2}} \frac{\mu^{\epsilon/2}g_2}{M_W}
      \left[\left(\begin{array}{ccc} \overline{u} & \overline{c} &
	      \overline{t} \end{array}\right)_R M_U V
	      \left(\begin{array}{c} d \\ s \\ b \end{array}\right)_L
	     -\left(\begin{array}{ccc} \overline{u} & \overline{c} &
	     \overline{t} \end{array}\right)_L V M_D
	    \left(\begin{array}{c} d \\ s \\ b \end{array}\right)_R
	     \right]\phi_+ \nonumber\\
	&  +&h.c..\label{2}
\end{eqnarray}
Where V represents the $3 \times 3$ unitary Kobayashi-Maskawa matrix,
$M_U$ and $M_D$ denote the diagonalized quark mass matrices, the
subscript
$L$ and $R$ denote left-handed and right-handed quarks, respectively.
This charged current, which is the main cause of flavor changing
neutral current, is pure left-handed.

Like ref.\cite{fuj}, a phenomenological coupling of the right-handed
top and bottom quarks to the W boson is introduced:
\begin{equation} {\cal L}= g_2/\sqrt{2}
\mu^{\epsilon/2} V_{tb} f_R^{tb} \overline{t}_R \gamma_{\mu} b_R
W_+^{\mu} +\frac{\mu^{\epsilon/2} g_2}{\sqrt{2} M_W}
V_{tb} f_R^{tb} \overline{t} ( m_t P_R -m_b P_L) b \phi_+ +h.c.,\label{3}
\end{equation}
with $f_R^{tb}$ denoting the strength of this additional coupling.
Using this interaction (\ref{3}) together with (\ref{2}), we performed
the QCD corrections to $b\to s\gamma$ decay.

At first, we integrate out the top quark, generating
an effective five quark theory, introducing dimension-5 and dimension-6
 effective operators as to include effects of the absent top quark.
Higher dimension operators are suppressed by factor of $p^2/m_t^2$, here
$p^2$ characterizes the interesting external momentum of b quark
$p^2\sim m_b^2$. For leading order of $m_b^2/m_t^2$, dimension-6
operators are good enough to make a complete basis of
operators\cite{Cho,lcd1}\footnote{Notice here $W_{LR}^2$ is a new
operator compared with SM case.}:
\begin{eqnarray}
dimension~ 5: &&\nonumber\\
O_{LR}^1  & =  &  -\frac{1}{16\pi^2} m_b \overline{s}_L D^2 b_R,
\nonumber\\
O_{LR}^2  &  =  &  \mu^{\epsilon/2} \frac{g_3}{16\pi^2}
	m_b \overline{s}_L \sigma^{\mu\nu} X^a b_R G_{\mu\nu}^a,
\nonumber\\
O_{LR}^3  &  =  &  \mu^{\epsilon/2} \frac{e Q_b}{16\pi^2}
	m_b \overline{s}_L \sigma^{\mu\nu} b_R	F_{\mu\nu},
\nonumber\\
Q_{LR}  &  =  &  \mu^{\epsilon} g_3^2 m_b
	\phi_{+}\phi_{-} \overline{s}_L b_R,
\nonumber\\
W_{LR}^1 &  =  & -i \mu^{\epsilon} g_3^2 m_b W^{\nu}_{+}W_{-}^{\mu}
	\overline{s}_L \sigma_{\mu \nu} b_R, \nonumber\\
W_{LR}^2 &  =  &  \mu^{\epsilon} g_3^2 m_b W^{\mu}_{+}W_{-}^{\mu}
	\overline{s}_L b_R,\nonumber\\
dimension ~6: &&\nonumber\\
P_L^{1,A}  &  =  &  -\frac{i}{16\pi^2}  \overline{s}_L
  T_{\mu\nu\sigma}^A D^{\mu} D^{\nu} D^{\sigma} b_L,\nonumber\\
P_L^2 & = & \mu^{\epsilon/2} \frac{e Q_b}{16\pi^2}  \overline{s}_L
	\gamma^{\mu} b_L \partial^{\nu} F_{\mu\nu},\nonumber\\
P_L^4  &  =  &  i \mu^{\epsilon/2} \frac{e Q_b}{16\pi^2}
	\tilde{F}_{\mu\nu}
	\overline{s}_L \gamma^{\mu} \gamma^5 D^{\nu} b_L,\nonumber\\
R_L^1 &  =  &  i \mu^{\epsilon} g_3^2 \phi_{+}\phi_{-} \overline{s}_L
	\not \!\! D b_L,\nonumber\\
R_L^2  &  =  &  i \mu^{\epsilon} g_3^2(D^{\sigma} \phi_+) \phi_{-}
	\overline{s}_L\gamma_{\sigma} b_L,
\nonumber\\
W_L^1  &  =  &  i \mu^{\epsilon} g_3^2 W^{\nu}_{+}W_{-}^{\mu}
	\overline{s}_L
	\gamma _{\mu}\not \!\! D \gamma _{\nu} b_L,\nonumber\\
W_L^2  &  =  &  i \mu^{\epsilon} g_3^2(D^{\sigma} W^{\nu}_+) W^{\mu}_{-}
	\overline{s}_L \gamma_{\mu} \gamma_{\sigma} \gamma_{\nu} b_L,
\nonumber\\
W_L^3  &  =  &  i \mu^{\epsilon} g_3^2 W_{+\mu} W^{\mu}_{-}
\overline{s}_L \stackrel{\leftrightarrow}{\not \!\! D} b_L, \nonumber\\
W_L^4  &  =  &  i \mu^{\epsilon} g_3^2 W^{\nu}_+ W^{\mu}_{-}
\overline{s}_L (\stackrel{\leftrightarrow}{D}_{\mu}\! \gamma_{\nu} +
	 \gamma_{\mu}\! \stackrel{\leftrightarrow}{D}_{\nu} ) b_L.
\end{eqnarray}
Where $\overline{s}_L\! \stackrel{\leftrightarrow}{D}_{\mu}
\!\gamma_{\nu} b_L$
stands for $(\overline{s}_L D_{\mu} \gamma_{\nu} b_L +(D_{\mu}
 \overline{s}_L)
 \gamma_{\nu} b_L)$ and the covariant derivative is defined as
$$D_{\mu}=\partial_{\mu}-i\mu^{\epsilon/2}g_3 X^a G_{\mu}^{a} -
i \mu^{\epsilon/2}eQ A_{\mu},$$
with $g_3$ denoting the QCD coupling constant.
The tensor $T_{\mu\nu\sigma}^A$ appearing in $P_L^{1,A}$
 assumes the following
Lorentz structure, the index $A$ ranging from 1 to 4:
\begin{equation}
\begin{array}{ll}
        T_{\mu\nu\sigma}^1=g_{\mu\nu} \gamma_{\sigma},
&T_{\mu\nu\sigma}^2=g_{\mu\sigma} \gamma_{\nu},\nonumber\\
	T_{\mu\nu\sigma}^3=g_{\nu\sigma} \gamma_{\mu},
&T_{\mu\nu\sigma}^4=-i \epsilon_{\mu\nu\sigma\tau}
	\gamma^{\tau} \gamma_5.
\end{array}
\end{equation}

In our following calculations, we try to work in a background field
$R_{\xi}$ gauge (with $\xi =1$)\cite{Abbott}, in order to maintain
explicit gauge invariance in calculations of anomalous dimensions.
Furthermore the usually trilinear interaction between photon,
W boson, and would-be Goldstone boson vanish in this gauge.
So that, we did not include operators which involve both W boson
and would-be Goldstone boson in one operator, like $W_+^{\nu} \phi
_- \overline{s}_L \gamma_{\nu} b_L$.
With the above operators, we can write down our effective Hamiltonian as
\begin{equation}
{\cal H}_{eff}=2 \sqrt{2} G_F V_{tb}V_{ts}^{\ast}\displaystyle \sum _i
C_i(\mu)O_i(\mu). \label{eff}
\end{equation}
The coefficients $C_i(m_t )$ of operators can be calculated from full
theory by matching conditions, keeping only leading orders of
$p^2/m_t^2$\cite{Cho,lcd1}.
Terms proportional to $f_R^{tb}m_t/m_b$ are from right handed current.
Since $m_t/m_b$ is a very large value, it is very convenient for us
to keep only leading orders of $m_t/m_b$. This is also the reason
why only top-bottom right-handed charge current is introduced, while
other up-down, charm-strange right-handed currents are ignored.
\begin{eqnarray}
C_{R_L^1} &=& C_{R_L^2}\;\;=\;\;1/g_3^2,\nonumber\\
C_{Q_{LR}}&=& -\left( 1- f_R^{tb} \frac{m_t}{m_b} \right) /g_3^2,\nonumber\\
C_{W_{LR}^1} &=& C_{W_{LR}^2} \;\;=\;\; \frac{m_t}{m_b}f_R^{tb}\frac{\delta}
	{g_3^2},\nonumber\\
C_{W_L^1} &=& C_{W_L^2}\;\;=\;\;\delta /g_3^2,\nonumber\\
C_{W_L^3} &=& C_{W_{L}^4}\;\;=0.\label{coe1}
\end{eqnarray}
The other coefficients of operators are all from the integrations of
electroweak loops. Terms like $\log(\mu^2/m_t^2)$ in coefficients
of operators $O_{LR}^3$, $P_L^2$ and $P_L^4$
vanish here, because of the matching scale $\mu=m_t$. They will
be regenerated at lower scales by renormalization group running
of electroweak later. The other logarithms are all from the finite part
integration of loops, for there are two different mass scale particles
in one loop.
\begin{eqnarray}
C_{O_{LR}^1}&=&-\left(\frac{1+\delta}
	{2(1-\delta)^2}+\frac{\delta}{(1-\delta)^3}\log\delta\right)
+f_R^{tb} \frac{m_t}{m_b} \left(\frac{1-3\delta-4\delta^2}
	{2(1-\delta)^2}+\frac{\delta-4\delta^2}{(1-\delta)^3}\log\delta\right),
\nonumber\\
C_{O_{LR}^2}&=&-\frac{1}{2} \left(\frac{1}{(1-\delta)}+
	\frac{\delta}{(1-\delta)^2}\log\delta\right)
\left( 1- f_R^{tb} \frac{m_t}{m_b} \right), \nonumber\\
C_{O_{LR}^3}&=& \left(\frac{1}{(1-\delta)}+
	\frac{\delta}{(1-\delta)^2}\log\delta\right)
+f_R^{tb} \frac{m_t}{m_b} \left(\frac{-1+6\delta}{(1-\delta)}+
	\frac{-\delta+12\delta^2}{(1-\delta)^2}\log\delta
	+6 \delta \log \frac{\mu^2}{m_t^2} \right), \nonumber\\
C_{P_L^{1,1}}&=& C_{P_L^{1,3}}\;\;=\;\;
	\left(\frac{\frac{11}{18}+\frac{5}{6}\delta
	-\frac{2}{3}\delta^2 +\frac{2}{9} \delta ^3}{(1-\delta)^3}+
\frac{\delta+\delta^2-\frac{5}{3}\delta^3 +\frac{2}{3} \delta ^4}
	{(1-\delta)^4}\log\delta\right),\nonumber\\
C_{P_L^{1,2}}&=&\left(\frac{-\frac{8}{9}-\frac{1}{6}\delta
	+\frac{17}{6}\delta^2 -\frac{7}{9} \delta ^3}{(1-\delta)^3}+
	\frac{-\delta+\frac{10}{3}\delta^3 -\frac{4}{3} \delta^4}
	{(1-\delta)^4}\log \delta \right),\nonumber\\
C_{P_L^{1,4}}&=&\left(\frac{\frac{1}{2}-\delta
	-\frac{1}{2}\delta^2 +\delta^3}{(1-\delta)^3}+
\frac{\delta-3\delta^2+2\delta^3}{(1-\delta)^4}\log\delta\right),\nonumber\\
C_{P_L^2}&=&\frac{1}{Q_b}\left(\frac{\frac{3}{4}+\frac{1}{2}\delta
	-\frac{7}{4}\delta^2 +\frac{1}{2} \delta^3 }{(1-\delta)^3}
	-\frac{1}{3} \delta
+\left(\frac{\frac{1}{6} +\frac{5}{6}\delta -\frac{5}{3}\delta^3+
	\frac{2}{3} \delta^4} {(1-\delta)^4}
	- \frac{1}{6} -\frac{1}{3} \delta \right)
	\log\delta \right)\nonumber\\
&&	-\frac{1}{2} \log \frac{\mu^2}{m_t^2} -\delta \log \frac{\mu^2}{m_t^2},
\label{coe}\\
C_{P_L^4}&=&\frac{1}{Q_b}\left(\frac{-\frac{1}{2}-5\delta
	+\frac{17}{2}\delta^2 -3\delta^3 }{(1-\delta)^3}+
	\frac{-5\delta +7\delta^2 -2\delta^3}{(1-\delta)^4}\log\delta
	+4\delta \log\delta \right)+12\delta \log \frac{\mu^2}{m_t^2},
\nonumber
\end{eqnarray}
where $\delta = M_W^2/m_t^2$.
When $f_R^{tb}=0$, the above results (\ref{coe1}), (\ref{coe})
reduce to that of SM case\cite{lcd1}.

The renormalization group equation satisfied by
the coefficient functions $C_i(\mu)$ is
\begin{equation}
\mu \frac{d}{d\mu} C_i(\mu)=\displaystyle\sum_{j}(\gamma^{\tau})_{
ij}C_j(\mu).\label{ren}
\end{equation}
The solution to this renormalization group equation (\ref{ren})
appears in obvious matrix notation as
\begin{equation}
C(\mu_2)=\left[\exp\int_{g_3(\mu_1)}^{g_3(\mu_2)}dg\frac
{\gamma^T(g)}{\beta(g)}\right] C(\mu_1),\label{solu}
\end{equation}
where the anomalous dimension matrix $\gamma_{ij}$
is calculated in practice by requiring
renormalization group equations for Green functions
with insertions of composite operators
to be satisfied order by order in perturbation theory\cite{Cho,lcd1}.
\begin{equation}
\begin{array}{rccl}
  \gamma=
   & \begin{array}{c} \\ Q_{LR}\\ R_L^1\\ R_L^2\\ W_{LR}^1\\ W_{LR}^2\\
	 W_L^1\\	W_L^2\\ W_L^3\\ W_L^4  \end{array}
   & \begin{array}{c}
  	\begin{array}{cccccccc}
O_{LR}^1& O_{LR}^2& O_{LR}^3 & P_L^{1,A} & P_L^2 & P_L^3 & P_L^4 &
  	\end{array} \\

	\left(\begin{array}{ccccccccccccc}
	 0 && 0 &&& 0 && 0 &  0   & 0 && 0 \\
	 0 && 0 &&& 0 && 0 &  0   & 0 && 0 \\
         0 && 0 &&& 0 && 0 & -1/2 & 0 && 0 \\
	 0 && 0 &&& 6 && 0 &  0   & 0 && 0\\
         0 && 0 &&& 0 && 0 &  0   & 0 && 0 \\
	 0 && 0 &&& 0 && 0 &  0   & 0 && 12\\
	 0 && 0 &&& 0 && 0 & -1   & 0 && 0 \\
	 0 && 0 &&& 0 && 0 &  0   & 0 && 0 \\
	 0 && 0 &&& 0 && 0 &  0   & 0 && 0 \\
	\end{array}\right)

     \end{array}

   & 16\pi^2\; \displaystyle{ \frac{g_3^2}{8\pi^2} }.

\end{array}\label{weak}
\end{equation}
These mixings are all between operators (Q,R,W) induced by tree-diagram and
operators (O,P) induced by loop-diagram. In order to see how the
renormalization
group method is accomplished, we neglect the proper QCD corrections for the
moment, so that we can take into account only the above entries of anomalous
dimensions. Insert this matrix to eqn.(\ref{solu}), we find the
following relations:
\begin{eqnarray*}
C_{O_{LR}^3} ( M_W) &=& C_{O_{LR}^3} (m_t) +6\frac{m_t}{m_b} f_R^{tb}
	\delta \log \frac{\mu^2}{m_t^2},\\
C_{P_{L}^2} ( M_W) &=& C_{P_{L}^2} (m_t) - \frac{1}{2} \log\frac{\mu^2}{m_t^2}
	-\delta \log \frac{\mu^2}{m_t^2},\\
C_{P_{L}^4} ( M_W) &=& C_{P_{L}^4} (m_t) + 12 \delta \log \frac{\mu^2}{m_t^2}.
\end{eqnarray*}

Here the renormalization group equation reproduces the
$\log(\mu^2/m_t^2)$ terms in the coefficients of operators at
equation (\ref{coe}) which vanished at $\mu=m_t$.
This proves the consistence of the whole calculation.

The QCD anomalous dimensions for each of the operators
in our basis are\cite{Cho,lcd1}:
\begin{equation}
 \begin{array}{c}
     \begin{array}{cccccccccccc}
	& && O_{LR}^1 & O_{LR}^2& O_{LR}^3& P_L^{1,1}& P_L^{1,2}&
	  P_L^{1,3}& P_L^{1,4}& P_L^{2}& P_L^{4}
     \end{array}\\
     \begin{array}{r}
  O_{LR}^1\\ O_{LR}^2\\ O_{LR}^3\\ P_L^{1,1}\\ \gamma=\; P_L^{1,2}\\
	  P_L^{1,3}\\ P_L^{1,4}\\ P_L^{2}\\ P_L^{4}
     \end{array}\left(\begin{array}{cccccccccccccc}
	  \frac{20}{3} && 1 && -2 & 0 & 0 & 0 & 0 && 0  && 0 \\
 -8 && \frac{2}{3} && \frac{4}{3} & 0 & 0 & 0 & 0 && 0 && 0 \\
	 0 && 0 && \frac{16}{3} & 0 & 0 & 0 & 0 && 0 && 0 \\
	 6 && 2 && -1 & \frac{2}{3} & 2 & -2 & -2 && 0 && 0 \\
	 4 && \frac{3}{2} && 0 & -\frac{113}{36} & \frac{137}{18}
	 & -\frac{113}{36} &-\frac{4}{3} &&\frac{9}{4} && 0  \\
	 2 && 1 && 1 & -2 & 2 & \frac{2}{3} & -2 && 0 && 0 \\
	 0 && \frac{1}{2} && 2 & -\frac{113}{36} & \frac{89}{18}
	  & -\frac{113}{36} &
	 \frac{4}{3} && \frac{9}{4} && 0 \\
	 0 && 0 && 0 & 0 & 0 & 0 & 0 && 0 && 0 \\
	 0 && 0 && -\frac{4}{3} & 0 & 0 & 0 & 0 && 0 && 0
	\end{array}\right) \displaystyle{ \frac{g_3^2}{8\pi^2} },
\end{array}
\label{anom1}
\end{equation}

\begin{equation}
\begin{array}{cccc}
\gamma= & \begin{array}{c}
	\\ Q_{LR}\\ R_L^1\\ R_L^2\\ W_{LR}^1\\ W_{LR}^2\\
		W_L^1\\ W_L^2\\	W_L^3\\ W_L^4\\
          \end{array}

	& \begin{array}{c}
	      \begin{array}{ccccccccc}Q_{LR} & R_L^1 & R_L^2
			& W_{LR}^1 & W_{LR}^2 & W_L^1 & W_L^2 & W_L^3 & W_L^4
 		     \end{array}
	    \\
	      \left( \begin{array}{ccccccccccccccc}
 \frac{23}{3} && 0  & 0 && 0 && 0 && 0 && 0 &  0  &      0\\
 0  && \frac{23}{3} & 0 && 0 && 0 && 0 && 0 &  0  &     0\\
 0  &&  0 &\frac{23}{3}&& 0 && 0 && 0 && 0 &  0  &    0\\
 0  &&  0   &     0     &&13 && 0 && 0 && 0  &  0  &     0\\
 0  &&  0   &   0   &&  0 && \frac{23}{3} && 0 && 0  &  0  &     0\\
 0  &&  0   &   0   && -\frac{8}{3} && 0 &&\frac{23}{3} && 0 &-\frac{8}{9}
&\frac{16}{9}\\
 0  &&  0   &   0  && 0 && 0 && 0 && \frac{23}{3} &  0  &    0\\
 0  &&  0   &   0  && 0 && 0 && 0 && 0  & \frac{23}{3} &    0\\
 0  &&  0   &   0  && 0 && 0 && 0 && 0 & -\frac{16}{9} & \frac{101}{9}\\
	      \end{array} \right)
	  \end{array}
	& \displaystyle{ \frac{g_3^2}{8 \pi^2} }.

\end{array}\label{anom2}
\end{equation}
Notice here only the anomalous dimensions relevant to $W_{LR}^2$ are
new to the SM case.
To proceed with such complicated anomalous dimensions with so many
operators, the algebra computation code REDUCE is used. Combining the
above three matrices (\ref{weak},
\ref{anom1},\ref{anom2}) into one large 18$\times$18 matrix, we diagonalize
it analytically to give 18 eigenvalues and 18 eigenvectors.
Using equation (\ref{solu}), one can have the coefficients of
operators at $\mu=M_W$. All analytic expressions of
coefficients at this scale, which are functions of coefficients at
$\mu=m_t$, are included in the appendix.

In order to continue running the basis operator coefficients down to
lower scales, one must integrate out the weak gauge bosons and
would-be Goldstone bosons at $\mu=M_W$ scale.
 From matching conditions, one finds the following relations
between coefficient functions just below(-) and above(+)
$\mu=M_W$\cite{Cho,lcd1}:
\begin{eqnarray}
C_{O_{LR}^1}(M_W^-)&=&C_{O_{LR}^1}(M_W^+), \nonumber\\
C_{O_{LR}^2}(M_W^-)&=&C_{O_{LR}^2}(M_W^+), \nonumber\\
C_{O_{LR}^3}(M_W^-)&=&C_{O_{LR}^3}(M_W^+),\nonumber \\
C_{P_L^{1,1}}(M_W^-)&=&C_{P_L^{1,1}}(M_W^+) + 2/9,\nonumber \\
C_{P_L^{1,2}}(M_W^-)&=&C_{P_L^{1,2}}(M_W^+) - 7/9, \nonumber\\
C_{P_L^{1,3}}(M_W^-)&=&C_{P_L^{1,3}}(M_W^+) + 2/9,\nonumber \\
C_{P_L^{1,4}}(M_W^-)&=&C_{P_L^{1,4}}(M_W^+) + 1, \nonumber\\
C_{P_L^2}(M_W^-)&=&C_{P_L^2}(M_W^+)
	-C_{W_L^2}(M_W^+) - 3/2, \nonumber\\
C_{P_L^3}(M_W^-)&=&C_{P_L^3}(M_W^+), \nonumber\\
C_{P_L^4}(M_W^-)&=&C_{P_L^4}(M_W^+) + 9.\label{14}
\end{eqnarray}
The operators involving Goldstone $\phi$ and W bosons are absent after
this matching.

In addition to these, there are new four-quark operators from
the matching\cite{Grin,Mis,Ciu}:
\begin{equation}
\begin{array}{ll}
O_1=(\overline{c}_{L\beta} \gamma^{\mu} b_{L\alpha})
	    (\overline{s}_{L\alpha} \gamma_{\mu} c_{L\beta}),
& O_2=(\overline{c}_{L\alpha} \gamma^{\mu} b_{L\alpha})
	    (\overline{s}_{L\beta} \gamma_{\mu} c_{L\beta}),\\
O_3=(\overline{s}_{L\alpha} \gamma^{\mu} b_{L\alpha})
	\displaystyle{\sum _q} (\overline{q}_{L\beta} \gamma_{\mu}
	q_{L\beta}),
& O_4=(\overline{s}_{L\alpha} \gamma^{\mu} b_{L\beta})
	\displaystyle{\sum _q} (\overline{q}_{L\beta} \gamma_{\mu}
	q_{L\alpha}),\\
O_5=(\overline{s}_{L\alpha} \gamma^{\mu} b_{L\alpha})
	 \displaystyle{\sum _q} (\overline{q}_{R\beta}
	\gamma_{\mu} q_{R\beta}),
& O_6=(\overline{s}_{L\alpha} \gamma^{\mu} b_{L\beta})
	\displaystyle{\sum _q} (\overline{q}_{R\beta}
	\gamma_{\mu} q_{R\alpha}),
\end{array}
\end{equation}
with coefficients
$$ C_i(M_W)=0, \;\; i=1,3,4,5,6, \;\; C_2(M_W)=1.$$
Although only coefficient of $O_2$ is not zero, the other operators
should also be included, for their mixing through renormalization group
running.

To simplify the calculation and compare with the previous results,
equations of motion(EOM)\cite{eom} is used to reduce all the remaining
two-quark operators to the gluon and photon magnetic moment
operators $O_{LR}^2$ and $O_{LR}^3$. Neglecting the strange quark
mass in comparison with the bottom quark
mass, we obtain the on-shell equivalence relations:
\begin{eqnarray}
O_{LR}^1&=& P_L^{1,1}~~=~~-\frac{1}{2}O_{LR}^2-\frac{1}{2}O_{LR}^3,\nonumber\\
P_L^{1,2}&=&-P_L^{1,4}~~=~~-\frac{1}{4}O_{LR}^2-\frac{1}{4}O_{LR}^3,\nonumber\\
P_L^{1,3}&=&P_L^2~~=~~ 0,\nonumber\\
P_L^4&=&-\frac{1}{4}O_{LR}^3.\label{rela}
\end{eqnarray}

To be comparable with previous results without QCD corrections from $m_{top}$
to $M_W$, operators $O_{LR}^3$, $O_{LR}^2$ are rewritten
as $O_7$, $O_8$ like ref.\cite{Grin,Mis,Ciu},
\begin{eqnarray}
O_7&=&(e/16\pi^2) m_b \overline{s}_L \sigma^{\mu\nu}
	    b_{R} F_{\mu\nu},\nonumber\\
O_8&=&(g/16\pi^2) m_b \overline{s}_{L} \sigma^{\mu\nu}
	    T^a b_{R} G_{\mu\nu}^a.
\end{eqnarray}
Then
the operator basis now consists of 8 operators. Using eqn.(\ref{14})
(\ref{rela}) together with (\ref{23}) in the appendix,
the  effective Hamiltonian appearing just below the W-scale is easily
drawn out:
\begin{eqnarray}
{\cal H}_{eff} && =\frac{4G_F}{\sqrt{2}} V_{tb}V_{ts}^{\ast}
	\displaystyle \sum_{i}
                C_i(M_W^-) O_i(M_W^-)\nonumber\\
	& &\stackrel{EOM}{\rightarrow}
		\frac{4G_F}{\sqrt{2}} V_{tb}V_{ts}^*\left\{
		\displaystyle{\sum_{i=1}^{6} }C_i (M_W^-)O_i + C_7(M_W^-) O_7
		+C_8(M_W^-) O_8 \right\}.
\end{eqnarray}
with
\nopagebreak[1]
\begin{eqnarray}
C_{O_8}(M_W^-) &= & \left( \frac{\alpha _s (m_t)} {\alpha _s (M_W)}
	\right) ^{ \frac{14}{23} } \left\{ \frac{1}{2}C_{O_{LR}^1}(m_t)
-C_{O_{LR}^2}(m_t) +\frac{1}{2}C_{P_L^{1,1}}(m_t) \right.\nonumber\\
&&	\;\;\;\;\;\;\;\;\;\;\;\;\;\;\;\;\;\;
	\left.+\frac{1}{4}C_{P_L^{1,2}}(m_t)
	-\frac{1}{4}C_{P_L^{1,4}}(m_t)\right\}
	-\frac{1}{3} ,\label{c2}
\end{eqnarray}
\begin{eqnarray}
&\displaystyle{
C_{O_7}(M_W^-) = \frac{1}{3}\left( \frac{\alpha _s (m_t)} {\alpha _s (M_W)}
	\right) ^{ \frac{16}{23} } \left\{ C_{O_{LR}^3}(m_t)
	+8 C_{O_{LR}^2}(m_t) \left[1-\left( \frac{\alpha _s (M_W)}
{\alpha _s (m_t)} \right) ^{ \frac{2}{23} } \right]\right.}&\nonumber\\
&\displaystyle{	+\left[-\frac{9}{2} C_{O_{LR}^1}(m_t)
	-\frac{9}{2}C_{P_L^{1,1}}(m_t)
-\frac{9}{4}C_{P_L^{1,2}}(m_t) +\frac{9}{4}C_{P_L^{1,4}}(m_t)\right]
\left[1- \frac{8}{9} \left( \frac{\alpha _s (M_W)}
{\alpha _s (m_t)} \right) ^{ \frac{2}{23} } \right] }&\nonumber\\
&\displaystyle{	\;\;\;\left.-\frac{1}{4}C_{P_L^4}(m_t)
+\frac{9}{23} 16\pi^2 C_{W_L^1}(m_t) \left[1- \frac{\alpha _s (m_t)}
	{\alpha _s (M_W)} \right] \right \}
	-\frac{23}{36} }. &  \label{c3}
\end{eqnarray}
Since they are expressed by coefficients of operators
at $\mu=m_t$ and QCD coupling $\alpha_s$,
it is convenient to utilize these formulas.

If QCD correction from $m_t$ to $M_W$ is neglected[by setting $\alpha_s(
M_W)=\alpha_s(m_t)$ in the above equations (\ref{c2}),(\ref{c3})], the
results are reduced exactly to those
the previous authors gave\cite{fuj,cho1}.

The QCD corrections from $\mu=M_W$ to $\mu=m_b$
have been a subject of many papers\cite{Grin,Grig,Cel,Mis,Yao,Ciu}.
It has attracted great interest recent years, since most QCD corrections
come from this stage. For a long time, there are some discrepancies in
this calculation, for it is rather lengthy calculation involving 2-loop
diagrams. Until recently, it is completely resolved, and a complete
leading log result is given\cite{Ciu}.
Here, we are free to utilize these results,
so the coefficients of operators at $\mu=m_b$ scale are
\begin{equation}
C_7^{eff}(m_b) = \eta^{16/23}C_7(M_W) +\frac{8}{3} ( \eta^{14/23}
-\eta^{16/23} ) C_8(M_W) +C_2(M_W) \displaystyle \sum _{i=1}^{8} h_i
\eta^{a_i}.
\end{equation}
With $\eta = \alpha_s(M_W) /\alpha_s (m_b)$,
$$ h_i =\left( \frac{626126}{272277},~ -\frac{56281}{51730}, ~
-\frac{3}{7},~ -\frac{1}{14},~ -0.6494,~ -0.0380,~ -0.0186,~ -0.0057 \right),$$
$$a_i = \left( \frac{14}{23}, ~\frac{16}{23}, ~\frac{6}{23},
{}~-\frac{12}{23} \right).$$
Here $m_b=4.9$GeV is used.

\section{The $\overline{B} \rightarrow X_s \gamma$ decay rate}

Following previous authors\cite{Grin,Mis,Ciu}, one obtains the
$\overline{B} \rightarrow X_s \gamma$ decay rate normalized to
the quite well established semileptonic decay rate $BR(\overline{B}
\rightarrow X_c e\overline{\nu})$.
\begin{equation}
BR(\overline{B} \rightarrow X_s \gamma) /BR(\overline{B}
\rightarrow X_c e\overline{\nu}) \simeq\Gamma(b\rightarrow
s\gamma)/\Gamma
(b\rightarrow ce\overline{\nu}).
\end{equation}
Then
\begin{equation}
\frac{BR(\overline{B} \rightarrow X_s \gamma)}{BR(\overline{B}
\rightarrow X_c e \overline{\nu})} \simeq \frac{6 \alpha_{QED}}{\pi
 g (m_c/m_b)}
|C_7^{eff}(m_b)|^2 \left(1-\frac{2 \alpha_{s}(m_b)}{3 \pi} f(m_c/m_b)
\right)^{-1},
\end{equation}
where $g(m_c/m_b)\simeq 0.45$ and $f(m_c/m_b)\simeq 2.4$ correspond
to the phase space
factor and the one-loop QCD correction to the semileptonic decay,
respectively\cite{Cabi}. The electromagnetic fine structure constant
evaluated at the $b$ quark scale takes value as $\alpha_{QED}(m_b)=
1/132.7$. If we take experimental result $Br(\overline{B} \to
X_c e\overline{\nu} ) =10.8\% $\cite{data}, the branching ratios of
$\overline{B} \to X_s \gamma$ is found. Compared with the previous
results without QCD running from $m_t$ to $M_W$\cite{fuj}, the
branching ratio is more sensitive to the value of $f_R^{tb}$. It is
no longer a pure enhancement like SM case. It is rather suppressed when
$|f_R^{tb}|$ takes large value, for example, When $|f_R^{tb}|=-0.08$,
the branching ratio is only about $1/10$ of that with QCD
correction from $m_t$ to $M_W$ neglected. While when $|f_R^{tb}|$
takes small value, the branching ratio is rather enhanced,
especially when $|f_R^{tb}|=0$, it is enhanced 11\% corresponding
to SM case. Since the right-handed current contribution is proportional
to $|f_R^{tb}|^2$, the whole QCD correction to $b\to s \gamma$ decay
is always a large enhancement.

In Fig 1, the branching ratio of $b\to s\gamma$ is displayed as
functions of $f_R^{tb}$ with $m_t=174$GeV\cite{CDF}. The three
lines correspond to $\alpha_s(m_Z)=0.107,~0.117,~0.127$. They are
all parabolic lines, because $Br(b\to s \gamma)$ is proportional to
$|C_7(m_b)|^2$, and $C_7(m_b)$ is proportional to $f_R^{tb}$.
That the three lines are very close to each other implies that
the branching ratios are not sensitive to the values of $\alpha_s$.

Fig.2 is branching ratios displayed as functions of $f_R^{tb}$, with
different top quark masses 158GeV, 174GeV and 190GeV. The QCD coupling
constant is $\alpha_s(m_Z)=0.117$. Since contributions from right-handed
current are proportional to $|f_R^{tb} m_t/m_b|^2$, the three
lines in Fig.2 diverge in the region far from $f_R^{tb} = 0$,
but nearly converge to a same line  when $f_R^{tb} \to 0$.

With the recent CLEO experiments of $b\to s\gamma$ decay, a new constraint
to right-handed current can be found. From Fig.1 and Fig.2, the parameter
$f_R^{tb}$ is constrained in two small windows $-0.087<f_R^{tb}< -0.050$
and $-0.023<f_R^{tb}<0.002$. Compared with results without QCD corrections
from $m_t$ to $M_W$\cite{fuj}, the whole line
transfers to left side(towards minus). So the windows also transfer
to minus side. Most allowed values of $f_R^{tb}$ are negative.

\section{Conclusion}

In the above, we have introduced an anomalous right-handed current
to top, bottom quarks and W boson coupling in the SM, and given the
full leading log QCD corrections(including
QCD running from $m_{top}$ to $M_W$) to $b\to s \gamma$ decay
in this model.

The QCD corrections from $m_{top}$ to $M_W$ to $b\to s \gamma$ decay
enhance the decay rate in small values of $f_R^{tb}$ like in SM case;
but suppress the decay rate when $f_R^{tb}$ takes larger value.
The whole QCD correction still makes a large enhancement.
The restrictions from $b\to s\gamma$ decay to anomalous top quark coupling
are strict, only two narrow windows are allowed. It is shown that the decay
$b\to s\gamma$ has been the most restrictive
process so far in constraining the parameters of the right-handed
current of top quark.

If a complete QCD next-leading log result of $b\to s\gamma$ decay is
performed, one can expect to obtain more precise results from $b\to
s \gamma$ decay by freely using our above results.

\section*{{Acknowledgement}}

The authors thank Prof. Z.M. Qiu, X.C Song, and Dr. Q.H. Zhang for
helpful discussions.

\appendix

\section{Operator coefficients at $\mu=M_W^+$ scale}

We quote here the results of the coefficients of operators at $\mu=M_W^+$
scale. The W bosons and would-be Goldstone bosons are not integrated out
yet.
\begin{eqnarray}
C_{O_{LR}^1}(M_W) &=& \left( 2 \zeta ^{14/23} -\zeta ^{8/23} \right)
	C_{O_{LR}^1}(m_t)
	+4 \left( - \zeta ^{14/23} +\zeta ^{8/23} \right) C_{O_{LR}^2}(m_t)
\nonumber\\
&&	+\left(2\zeta^{14/23} -\zeta^{8/23} -\zeta^{-4/23} \right)
	C_{P_L^{1,1}}(m_t)
\nonumber\\
&&	+\left(\zeta^{14/23} -\frac{89}{130}\zeta^{8/23} -\frac{113}{274}
	 \zeta^{-4/23} +\frac{864}{8905}\zeta^{113/138} \right)
	C_{P_L^{1,2}}(m_t)\nonumber\\
&&	+\left(-\zeta^{14/23} +\frac{171}{130}\zeta^{8/23} -\frac{113}{274}
	\zeta^{-4/23} +\frac{864}{8905}\zeta^{113/138} \right)
	 C_{P_L^{1,4}}(m_t),
\nonumber\\
C_{O_{LR}^2}(M_W) &=& \frac{1}{2} \left(\zeta^{14/23} -\zeta^{8/23} \right)
	  C_{O_{LR}^1}(m_t)
	+\left( -\zeta ^{14/23} +2\zeta ^{8/23} \right) C_{O_{LR}^2}(m_t)
\nonumber\\
&&	+\frac{1}{2} \left(\zeta^{14/23} -\zeta^{-4/23} \right)
	C_{P_L^{1,1}}(m_t)\nonumber\\
&&	+\left(\frac{1}{4} \zeta^{14/23} -\frac{6}{65}\zeta^{8/23}
	  -\frac{113}{548} \zeta^{-4/23} +\frac{432}{8905}\zeta^{113/138}
		\right) C_{P_L^{1,2}}(m_t)
\nonumber\\
&&	+\left(-\frac{1}{4} \zeta^{14/23} +\frac{53}{130}\zeta^{8/23}
	  -\frac{113}{548} \zeta^{-4/23} +\frac{432}{8905}\zeta^{113/138}
		\right) C_{P_L^{1,4}}(m_t),
\nonumber\\
C_{O_{LR}^3} (M_W) &=& \left( 5\zeta^{14/23} -\frac{1}{2} \zeta^{8/23}
	   -\frac{9}{2}\zeta^{16/23} \right) C_{O_{LR}^1}(m_t)
	+\left(-10\zeta^{14/23} +2\zeta^{8/23} +8\zeta^{16/23} \right)
	  C_{O_{LR}^2}(m_t)
\nonumber\\
&&	+\zeta^{16/23} C_{O_{LR}^3}(m_t)
	+\left(5 \zeta^{14/23} -\frac{1}{2} \zeta^{-4/23}
	  -\frac{9}{2} \zeta^{ 16/23} \right) C_{P_L^{1,1}}(m_t)
\nonumber\\
&&	+\left(\frac{5}{2} \zeta^{14/23} -\frac{6}{65}\zeta^{8/23}
	  -\frac{113}{548} \zeta^{-4/23} +\frac{432}{8905}\zeta^{113/138}
	   -\frac{9}{4} \zeta^{16/23}	\right) C_{P_L^{1,2}}(m_t)
\nonumber\\
&&	+\left(-\frac{5}{2} \zeta^{14/23} +\frac{53}{130}\zeta^{8/23}
	  -\frac{113}{548} \zeta^{-4/23} +\frac{432}{8905}\zeta^{113/138}
	   +\frac{9}{4} \zeta^{16/23}	\right) C_{P_L^{1,4}}(m_t)
\nonumber\\
&&	+\frac{1}{4} \left(1 -\zeta^{16/23} \right) C_{P_L^{4}}(m_t)
	+\frac{18}{23} \left(1- \zeta^{-1} \right) 16\pi^2 C_{W_{LR}^1}(M_W),
\nonumber\\
C_{P_L^{1,1}}(M_W)  &=& C_{P_L^{1,3}} \;\; =\;\; \zeta^{-4/23}
C_{P_L^{1,1}}(m_t)
	+\frac{113}{274} \left(\zeta^{-4/23} -\zeta^{113/138} \right)
	\left( C_{P_L^{1,1}}(m_t) +C_{P_L^{1,4}}(m_t) \right),
\nonumber\\
C_{P_L^{1,2}}(M_W)  &=& \left( \zeta^{8/23} -\zeta ^{-4/23} \right)
C_{P_L^{1,1}}(m_t)
	+ \left( \frac{1}{2} \zeta ^{8/23} -\frac{113}{274}\zeta^{-4/23}
	   +\frac{125}{137} \zeta^{113/138} \right) C_{P_L^{1,2}}(m_t)
\nonumber\\
&&	+\left(-\frac{1}{2} \zeta^{8/23} -\frac{113}{274} \zeta^{-4/23}
	  +\frac{125}{137} \zeta^{113/138} \right) C_{P_L^{1,4}}(m_t),
\nonumber\\
C_{P_L^{1,4}}(M_W)  &=& \left( -\zeta^{8/23} +\zeta ^{-4/23} \right)
	   C_{P_L^{1,1}}(m_t)
	+ \left( -\frac{1}{2} \zeta ^{8/23} +\frac{113}{274}\zeta^{-4/23}
	   +\frac{12}{137} \zeta^{113/138} \right) C_{P_L^{1,2}}(m_t)
\nonumber\\
&&	+\left(\frac{1}{2} \zeta^{8/23} +\frac{113}{274} \zeta^{-4/23}
	  +\frac{12}{137} \zeta^{113/138} \right) C_{P_L^{1,4}}(m_t),
\nonumber\\
C_{P_L^{2}} (M_W) &=& \frac{81}{226} \left( \zeta^{113/138} -1 \right)
	  \left( C_{P_L^{1,2}}(m_t) +C_{P_L^{1,4}}(m_t) \right )
	+ C_{P_L^{2}}(m_t)
\nonumber\\
&&	+\frac{3}{46} (1- \zeta ) (2\delta +1) 16\pi^2/g_3^2(m_t),
\nonumber\\
C_{P_L^{4}} (M_W) &=& C_{P_L^{4}}(m_t) +\frac{36}{23} (\zeta -1 )
	16\pi^2 \delta/g_3^2(m_t),\nonumber\\
C_{W_{LR}^1}(M_W)  &=& \zeta^{ 39/23} C_{W_{LR}^1} (m_t) + \frac{1}{2}\left(
	\zeta-\zeta^{39/23} \right) \delta /g_3^2(m_t),
\nonumber\\
C_{W_L^3}(M_W) &=& \zeta \left( C_{W_L^3} (m_t) +\frac{1}{4} (\zeta
^{32/69} -1) (C_{W_L^1} (m_t) +2 C_{W_L^4}(m_t) )\right ),
\nonumber\\
C_{W_L^4}(M_W) &=& \zeta^{101/69} \left( C_{W_L^4} (m_t) +\frac{1}{2}
(1- \zeta^{-32/69} ) C_{W_L^1} (m_t) \right ),
\nonumber\\
C_{O_i} (M_W) &=& \zeta C_{O_i}(m_t), ~~\;\;\; O_i = Q_{LR},~ R_L^1,
	~ R_L^2, ~W_{LR}^2, ~W_L^1, ~W_L^2.\label{23}
\end{eqnarray}
Here $\zeta =\alpha_s(m_t)/\alpha_s(M_W)$. The factor of $16\pi^2$ in
some of these expressions arises from mixing between operators induced
by tree diagrams and those by loop diagrams.


\section*{Figure Captions}
\parindent=0pt

Fig.1 BR($\overline{B} \rightarrow X_s \gamma$)
	as function of $f_R^{tb}$ for different $\alpha_s$ values.
$\alpha_s(m_Z) =0.107,~0.117,~0.127$.

Fig.2 BR($\overline{B} \rightarrow X_s \gamma$)
	as function of $f_R^{tb}$ for different top quark masses.
$m_t=158,~174,~190$GeV.

\newpage

\begin{figure}
\epsfbox{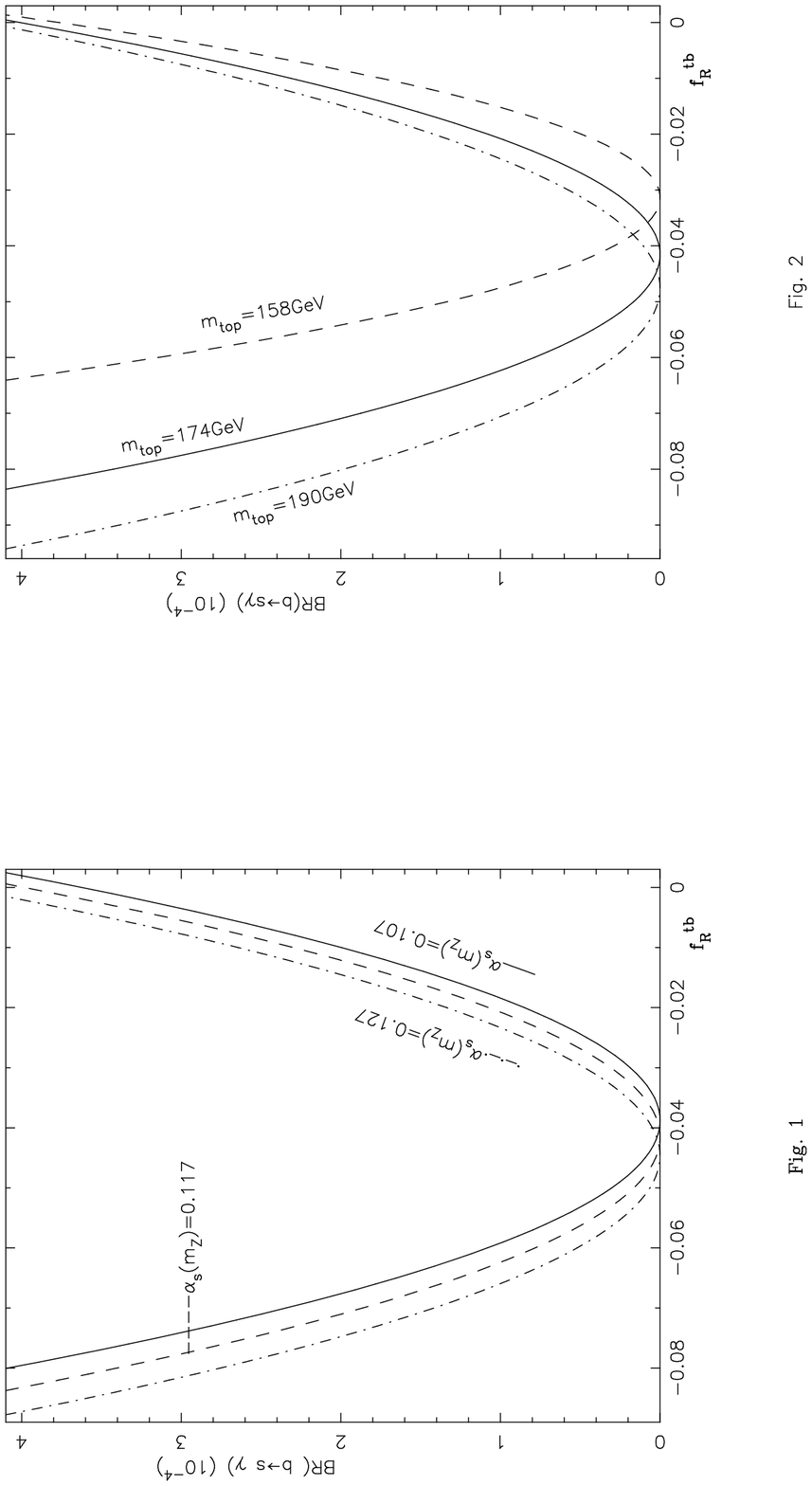}
\end{figure}


\begin{thebibliography}{88}

\bibitem{fuj} K. Fujikawa and A. Yamada, Phys. Rev. {\bf D49} (1994) 5890.
\bibitem{lr} J.C. Pati and A. Salam, Phys. Rev. {\bf D10} (1974) 275;
	R.N. Mohapatra and J.C. Pati, Phys. Rev. {\bf D11} (1975) 566, 2558;
	G. Senjanovic and R.N. Mohapatra, Phys. Rev. {\bf D12} (1975) 1502.
\bibitem{coc} D. Cocolicchio, G. Costa, G.L. Fogli, J.H. Kim and A. Masiero,
Phys. Rev {\bf D40} (1989) 1477.
\bibitem{cho1} P. Cho and M. Misiak, Phys. Rev. {\bf D49} (1994) 5894.
\bibitem{bab} K.S. Babu, K. Fujikawa and A. Yamada, Phys. Lett. {\bf B333}
(1994) 196.

\bibitem{Hew1} J.L. Hewett, SLAC preprint, SLAC-PUB-6521, 1994;
and references therein.
\bibitem{cleo2} E. Thorndike, CLEO Collaboration, talk given at the
{\it 27th International Conference on High Energy Physics}, Glasgow,
 1994.

\bibitem{cleo1} E. Thorndike, CLEO Collaboration, talk given at the
{\it 1993 Meeting of the American Physical Society}, Washington, D.C.,
April, 1993.


\bibitem{Grin} B. Grinstein, R. Springer and M.B. Wise, Phys. Lett.
	{\bf B202} (1988) 138; Nucl. Phys. {\bf B339} (1990) 269.
\bibitem{Grig} R. Grigjanis, P.J. O'Donnell, M. Sutherland, H, Navelet,
Phys. Lett. {\bf B213} (1988) 355; ibid. {\bf B286} 413(E).
\bibitem{Cel} G. Cella, G. Curci, G. Ricciardi, A. Vicere, Phys. Lett.
{\bf B248} (1990) 181; ibid. {\bf B325} (1994) 227.

\bibitem{Mis} M. Misiak, Phys. Lett. {\bf B269} (1991) 161;
Nucl. Phys. {\bf B393} (1993) 23.
\bibitem{Yao} K. Adel, Y.P. Yao, Mod. Phys. Lett. {\bf A8} (1993) 1679;
Phys. Rev. {\bf D49} (1994) 4945.
\bibitem{Ciu} M. Ciuchini, E. Franco, G. Martinelli, L. Reina,
 L. Silvestrini, Phys. Lett. {\bf B316} (1993) 127; M. Ciuchini,
 E. Franco, L. Reina, L. Silvestrini, Nucl. Phys. {\bf B421} (1994) 41.

\bibitem{Bur} A.J. Buras, M. Misiak, M. M\"unz and S. Pokorski,
	Nucl. Phys. {\bf B424} (1994) 374.

\bibitem{Cho} P. Cho, B. Grinstein, Nucl. Phys. {\bf B365} (1991) 279;
Erratum, {\bf B427} (1994) 697.
\bibitem{lcd1} C.S. Gao, J.L. Hu, C.D. L\"{u}, Z.M. Qiu, preprint
CCAST 93-28, hep-ph/9408351; C.S. Gao, J.L. Hu, C.D. L\"{u}, preprint
AS-ITP 94-45, hep-ph/9409258, to appear in Commun. Theor. Phys.

\bibitem{lcd2} C.D. L\"u, preprint, AS-ITP 94-32,
hep-ph/9408297, to appear in Nucl. Phys. B.
\bibitem{Anl} H. Anlauf, Nucl. Phys. {\bf B430} (1994) 245.

\bibitem{Abbott} L. Abbott, Nucl. Phys. {\bf B185} (1981) 189.
\bibitem{eom} H.D. Politzer, Nucl. Phys. {\bf B172} (1980) 349;
	H. Simma, preprint, DESY 93-083.

\bibitem{data} Particle Data Group, Phys. Rev. {\bf D45} (1992) No.11.

\bibitem{Cabi} N. Cabibbo and L. Maiani, Phys. Lett. {\bf B79}
 (1978) 109.

\bibitem{CDF} F. Abe, et al. CDF Collaboration, Phys. Rev. Lett. {\bf 73}
(1994) 225.

\end{thebibliography}
\end{document}